\documentclass[aps,prb,reprint,twocolumn,showpacs,floatfix,superscriptaddress,nofootinbib]{revtex4}
\usepackage{graphicx}
\usepackage{amssymb}
\usepackage{amsmath}
\usepackage{lmodern}
\usepackage{color}
\usepackage{hyperref}
\usepackage{empheq}
\usepackage[makeroom]{cancel}
\usepackage{epstopdf}
\begin{document}

\title{Poor man's scaling and Lie algebras}

\author{Eugene Kogan}
\email{Eugene.Kogan@biu.ac.il}
\affiliation{Jack and Pearl Resnick Institute, Department of Physics, Bar-Ilan University, Ramat-Gan 52900, Israel}
\affiliation{Max-Planck-Institut fur Physik komplexer Systeme,  Dresden 01187, Germany}
\affiliation{Donostia International Physics Center (DIPC), Paseo de Manuel Lardizabal 4, E-20018 San Sebastian/Donostia, Spain}

\begin{abstract}
We consider a general model, describing   a  quantum impurity with degenerate energy levels,  interacting with a gas of itinerant electrons,
derive general scaling equation for the model,
and analyse the connection between its particular forms  and the symmetry of interaction.
On the basis of this analysis
we write down  scaling equations for the
Hamiltonians which are the direct products of $su(3)$ Lie algebras and   have either
$SU(2)\times U(1)$ or $SU(2)$ symmetry. We also put into a new context  anisotropic Coqblin -- Schrieffer models proposed by us earlier.

\end{abstract}

%\pacs{75.50.Mm, 72.15.Qm, 03.75.Mn}

\maketitle

\section{Introduction}

In a seminal paper, published in 1964 and entitled
"Resistance Minimum in Dilute Magnetic Alloy" \cite{kondo},   Kondo considered
a (deceptively) simple model: magnetic impurity in a normal metal.
His, and the following theoretical analysis of the problem by different authors,
led to the appearance of many approaches and techniques, which became
 paradigms in  different fields of physics \cite{hewson}.
One of such approaches was the so called poor man's scaling, pioneered by Anderson \cite{anderson}.
 Models, similar to the one mentioned above, describe
magnetic ions in
a crystalline electric field \cite{noz,kroha2},
 tunnelling centres \cite{cox} and  system of quantum dots \cite{kroha,koerting,kikoin,krones,florens,delft,zalom}.
It is known that the  anisotropy can substantially change the physics of the model in comparison with the isotropic case
 \cite{shiba,yosida,irkhin2, costi,thomas,kogan,kogan2,kopietz}.

In our previous paper  \cite{kogan2} we considered a  {\it general} anisotropic model, describing a single quantum impurity with degenerate energy levels, interacting with a gas of itinerant electrons and derived poor man's scaling equation to the second order of interaction for the model.
We also introduced and studied the renormalizable anisotropic generalization of the Coqblin--Schrieffer (CS) model  \cite{coqblin}.
The CS model, though being well studied previously \cite{hewson,rajan,schlottmann,andrei,zlatic,bazhanov},  draw a lot of attention recently in connection with the   studies of quantum dots \cite{kikoin},  heavy fermions \cite{desgranges} and ultra-cold gases \cite{figueira,avishai}.

In the present contribution the algebraic structure of the general scaling equation is analyzed in a more detailed way.
We show the connection between the renormalizability of a Hamiltonian and the structure of the Lie algebra which is used to write down the Hamiltonian.
On the basis of this analysis we introduce still another renormalizable generalization of the CS model.

The rest of the paper is constructed as follows. In  Section \ref{se} we rederive the scaling equation and consider the case of the maximally symmetric interaction.
In  Section \ref{sym}  we  formulate the specific algebraic problem which is of interest to us, and, as the nontrivial example of its solution, derive the scaling equation for the case, when the operator of interaction acts on  $su(3)$ algebra, but has the symmetry  described either by $SU(2)\times U(1)$ or by $SU(2)$   subgroup of $SU(3)$.  In Section \ref{disc} we put  the anisotropic generalizations of the  CS model ($XXZ$ and $XYZ$ CS models) proposed by us earlier into the new context.  In Section \ref{combi} we try to combine the results of the previous two Sections. We conclude in Section \ref{conclusions}. Some mathematical details  are relegated to the Appendix.

\section{Scaling equation}
\label{se}

\subsection{Scaling equation and Lie  algebras}

The  Hamiltonian we start from is \cite{cox}
\begin{eqnarray}
\label{hamilto}
H=\sum_{{\bf k}\alpha}\epsilon_{\bf k}c_{{\bf k}\alpha}^{\dagger}c_{{\bf k}\alpha}
+\sum_{\substack{{\bf k},{\bf k}'\\\alpha\beta,ab}}V_{\beta\alpha,ba}X_{ba}c_{{\bf k}'\beta}^{\dagger}c_{{\bf k}\alpha},
\end{eqnarray}
where $c^{\dagger}_{{\bf k}\alpha}$ and $c_{{\bf k}\alpha}$ are  electron creation and annihilation operators of itinerant electron with  wave vector ${\bf k}$ and internal quantum number $\alpha$,   $\epsilon_{\bf k}$ is the energy of the electron; $X_{ba}=|b><a|$, where $|a>,|b>$ are the internal states of the scattering system, is the Hubbard $X$-operator.

We are interested in the low energy physics, that is in the electron states in the vicinity of the Fermi energy. However we can not just discard the   energy states of the electrons at the band edges, because virtual transitions to (from) these states influence the low energy physics.
The  idea of renormalization \cite{anderson} consists in decreasing the band width  of the itinerant electrons
from $[-D_0,D_0]$ to $[-D_0+|dD|,D_0-|dD|]$ ($dD<0$)
 and taking into account the terms which corresponded to virtual transitions through the electron states in the discarded energy intervals by renormalizing matrix elements of the perturbation connecting the states in the reduced band; this process is repeated many times, thus making the band width a running parameter.
In the lowest order of perturbation theory (one loop approximation) we obtain
 scaling equation
\begin{eqnarray}
\label{scaling}
\frac{dV_{\beta\alpha,ba}}{d\ln\Lambda}=\rho\sum_{\gamma,c}\left[V_{\beta\gamma,bc}V_{\gamma\alpha,ca}
-V_{\gamma\alpha,bc}V_{\beta\gamma,ca}\right],
\end{eqnarray}
where  $\rho$ is the density of states of itinerant electrons (assumed to be constant), which we further on will take being equal to 1
(more exactly,  $\rho$  can be eliminated by measuring all electronic densities in units of this
constant $\rho$); $\Lambda=D/D_0$.

Let the matrix $V_{\beta\alpha,ba}$ be
a sum of direct products of Hermitian matrices, acting in $ab$ and $\alpha\beta$ spaces respectively
\begin{eqnarray}
\label{product}
V=2\sum_{pp'}c_{pp'}G^p\otimes \Gamma^{p'},
\end{eqnarray}
where the set of matrices $\{G^p\}$ is closed with respect to commutation and hence generates some Lie algebra $g$. We assume that the set of matrices $\{\Gamma^p\}$ generates the algebra which is isomorphic to $g$.
(The indices $a,b$ of the matrix $G^p$ and the indices $\alpha,\beta$ of the matrix $\Gamma^p$ further on will be suppressed for brevity.)
Actually further on we will not distinguish these two algebras.
(The multiplier 2 is added to avoid the appearance of the multiplier $1/2$ later.)
%The matrix $c_{p\pi}$ is real, because the interaction should be Hermitian.

With the help of Eq. (\ref{product}) we can write down Eq. (\ref{scaling})  in a more transparent form
\begin{eqnarray}
\label{scaling25}
\sum_{pp'}\frac{dc_{pp'}}{d\ln\Lambda}G^p\otimes \Gamma^{p'}
=\sum_{ss'tt'}\left[G^s,G^t\right]\otimes \left[\Gamma^{s'},\Gamma^{t'}\right]c_{ss'}c_{tt'}.\nonumber\\
\end{eqnarray}
Introducing structure constants of the
 Lie algebras $g$
\begin{eqnarray}
\label{structure}
\left[G^s,G^t\right]=i\sum_pf^p_{st}G^p,\;\;\;
\left[\Gamma^{s},\Gamma^{t}\right]=i\sum_{p}f^{p}_{st}\Gamma^{p},
\end{eqnarray}
we can write down Eq. (\ref{scaling})  in an even more transparent form \cite{kogan2}
\begin{eqnarray}
\label{scaling2}
\frac{dc_{pp'}}{d\ln\Lambda}=-\sum_{sts't'}f^p_{st}f^{p'}_{s't'}c_{ss'}c_{tt'}.
\end{eqnarray}
From the point
of view of calculus, Eq.  (\ref{scaling2}) is (one of) the simplest scaling equation, one can consider.
However, it has an interesting  {\it algebraic} structure, which is the subject of the present communication.

\subsection{Maximally symmetric interaction}
\label{max}

The algebra $g$ defines the group $G$. If the symmetry of the interaction  is maximal (corresponds to the group $G$),
it (the interaction) is proportional  to the quadratic invariant of the group \cite{hall}
\begin{eqnarray}
\label{productc}
V=2J\sum_{p}G^p\otimes \Gamma^{p}.
\end{eqnarray}
Notice that   the Hamiltonian of free electrons has a maximal symmetry (in the electron internal quantum number space it is proportional to unit matrix), so we'll interchangeably talk about the symmetry of the Hamiltonian and the symmetry of the interaction.

Substituting $c_{ss'}=J\delta_{ss'}$ into the r.h.s. of Eq. (\ref{scaling2}) we obtain
\begin{eqnarray}
\label{scaling3}
\frac{dc_{pp'}}{d\ln\Lambda}=-J^2\sum_{st}f^p_{st}f^{p'}_{st}.
\end{eqnarray}
Assuming that the generators are self-dual
\begin{eqnarray}
\text{Tr}\left(G^pG^{p'}\right)=\frac{1}{2}\delta_{p,p'}
\end{eqnarray}
(same for $\Gamma$s),
we have
\begin{eqnarray}
\label{sc}
\sum_{st}f^p_{st}f^{p'}_{st}=N \delta_{p,p'},
\end{eqnarray}
where $N$ is the dimensionality of the matrices $G^p$.    \cite{georgi}
Thus the interaction (\ref{productc}) is renormalizable, and the scaling equation for such an interaction is
\begin{eqnarray}
\label{scaling4}
\frac{dJ}{d\ln\Lambda}=-N J^2.
\end{eqnarray}
(The results of this Subsection, being written down in mutually dual bases, one can find in Appendix \ref{curved}.)

Note that if we take $g=su(2)$, the  generators   being $\left\{G^p\right\}=\left\{S^x,S^y,S^z\right\}$ and $\left\{\Gamma^p\right\}=\left\{\frac{1}{2}\sigma^x,\frac{1}{2}\sigma^y,\frac{1}{2}\sigma^z\right\}$, the maximally ($SU(2)$) symmetric interaction is
\begin{eqnarray}
\label{productd}
V=J{\bf S}\cdot\boldsymbol\sigma,
\end{eqnarray}
which corresponds to
the isotropic
 Kondo model.
And if we take $g=su(N)$ and return from the generators notation to Hubbard operators and electron creation and annihilation operators notation,  the maximally ($SU(N)$) symmetric interaction is
\begin{eqnarray}
\label{ABCc}
V=J\sum_{mm'}^NX_{mm'}a_{m'}^{\dagger}a_{m}-\frac{1}{N}J\sum_{mm'}^NX_{mm}a_{m'}^{\dagger}a_{m'},
\end{eqnarray}
which corresponds to the CS  model\cite{coqblin}.

\subsection{Reduction to principal axes}
\label{su}

If the generators are chosen to be Hermitian,  the matrix $c_{pp'}$ is real.
If the matrix is in addition symmetric,
it  can be diagonalized by an orthogonal transformation of the generators.
(In Appendix \ref{DM} we consider a particular model when the matrix $C_{pp'}$ is not symmetric.)
Note that an orthogonal transformation (even a more general unitary one) does not change the commutation relations among the generators.
Thus
Eq. (\ref{product}) can be "reduced to the principal axes", that is to the form
\begin{eqnarray}
\label{pro}
V=2\sum_{p}J_pG^p\otimes \Gamma^p.
\end{eqnarray}

By substituting $J_{ss'}=J_s\delta_{s,s'}$ into the r.h.s. of Eq. (\ref{scaling2}) we obtain
\begin{eqnarray}
\label{scaling7}
\frac{dc_{pp'}}{d\ln\Lambda}=- \sum_{st}f^p_{st}f^{p'}_{st}J_sJ_t.
\end{eqnarray}
If
\begin{eqnarray}
\label{scaling77}
\sum_{st}f^p_{st}f^{p'}_{st}J_sJ_t=\delta_{p,p'}\sum_{st}\left(f_{st}^{p}\right)^2J_sJ_t,
\end{eqnarray}
the interaction (\ref{pro}) is renormalizable.

The condition (\ref{scaling77}) is fulfilled for the algebra $su(2)$, due to the fact that
for a given pair  $s,t$ the structure constant $f^p_{st}$ is different from zero for only one value of $p$, that is the sum in the r.h.s. of Eq. (\ref{structure}) consists of only a single summand. Thus the general interaction can be presented as
\begin{eqnarray}
\label{ab}
V=J_xS^x\otimes \sigma^x+J_yS^y\otimes \sigma^y+J_zS^z\otimes \sigma^z.
\end{eqnarray}
We call the model with the interaction (\ref{ab}) the $XYZ$ Kondo model

Studying commutation relations for  the algebra $su(3)$ presented in Appendix \ref{su21} shows that
for most pairs $s,t$, $f^p_{st}$ is different from zero only for one value of $p$. The pairs $4,5$ and $6,7$ result in two summands.
However, if we impose constraint
 $J_4J_5=J_6J_7$, Eq. (\ref{scaling77}) is still valid. We will assume that the constraint is imposed further on.
Thus  both for the $su(2)$ and for the $su(3)$ algebras the scaling equation becomes
\begin{eqnarray}
\label{scaling22}
\frac{dJ_{p}}{d\ln\Lambda}=-\sum_{st}\left(f^p_{st}\right)^2J_{s}J_{t}.
\end{eqnarray}
Further on in this paper  Eqs. (\ref{pro}) and (\ref{scaling22}) will be our starting point.

For the Kondo model (taking into account
the universally known $su(2)$ algebra commutation relations),  Eq. (\ref{scaling22}) takes the form of the system of three equations
\begin{eqnarray}
\label{scaling222}
\frac{dJ_i}{d\ln\Lambda}=-2J_jJ_k,
\end{eqnarray}
where $\{i,j,k\}$ is an arbitrary permutation of the Cartesian indices $x,y,z$

\section{Symmetry as the guiding principle}
\label{sym}

\subsection{A good symmetry is a (partially) broken symmetry}
\label{guide}

Following the previous Section, let us ask ourselves  what are the other particular {\it renormalizable}  cases of the general interaction
(\ref{pro}) (or, even more general, (\ref{product})), that is what are the constraints imposed on the interaction, which are respected by the renormalization? We  suggest symmetry as a guiding principle to find the answers.

A (too) simple case of the maximally symmetric interaction considered above corresponded to  the symmetry group of the Hamiltonian $G$ being  the group defined by the
algebra on which it acts. Let us now  decrease  the symmetry of the interaction from the group $G$ to some  its  subgroup $G'$.
It is obvious that the   general  interaction having this symmetry will be renormalizable.
It is also obvious that such an interaction will be an arbitrary linear combination of  all the invariant quadratic elements of the subgroup adjacently acting on the  algebra of the group.

To  illustrate the approach to renormalizability based on symmetry principle,
let us return to  the Kondo model and assume that the  symmetry of the interaction is $U(1)$; we take $S^z$ as the generator of the subgroup.
Due to decrease of symmetry, in addition to the existing  for $SU(2)$ symmetry invariant element $S^x\otimes \sigma^x+S^y\otimes \sigma^y+S^z\otimes \sigma^z$, there appears a new invariant element: $S^z\otimes \sigma^z$. For our purposes it is better to say that we have two invariants of the group $U(1)$:  $S^x\otimes \sigma^x+S^y\otimes \sigma^y$ and $S^z\otimes \sigma^z$.
Thus the most general interaction with the symmetry $U(1)$ can be written as an arbitrary linear combination of these invariant elements
\begin{eqnarray}
\label{aa}
V=J_x(S^x\otimes \sigma^x+S^y\otimes \sigma^y)+J_zS^z\otimes \sigma^z.
\end{eqnarray}
We call the model with the interaction (\ref{aa}) the $XXZ$ Kondo model.
For the interaction (\ref{aa}) the scaling equation is \cite{anderson}
\begin{eqnarray}
\label{scaling44}
\frac{dJ_x}{d\ln\Lambda}&=&- 2J_xJ_z\nonumber\\
\frac{dJ_z}{d\ln\Lambda}&=&- 2J_x^2.
\end{eqnarray}

Notice that $T^3$ generates Cartan subalgebra of the $su(2)$ algebra. This statement, though seeming simultaneously trivial and not very relevant,
is actually the seed out of which  Section \ref{disc} will grow from.

Further on we consider several  less simple cases, when we can explicitly realize the program, formulated above.
The  scaling equations we obtain,
to the best of our knowledge were not written down before. Everywhere, apart from the last Subsection,
$g=su(3)$.

\subsection{$U(2)$ ($SU(2)$) symmetry of the interaction}
\label{suu}

Let the symmetry of the interaction is either $SU(2)\times U(1)=U(2)$ or $SU(2)$.  In any case we have three invariant elements  (see Appendix \ref{su21}).  Thus we can write down the most general interaction with the prescribed symmetry as
\begin{eqnarray}
\label{ABC}
V=2J_A\sum_{p=1}^3G^p\otimes \Gamma^p+2J_B\sum_{p=4}^7G^p\otimes \Gamma^p+2J_CG^8\otimes \Gamma^8,\nonumber\\
\end{eqnarray}
where the generators $G^p$ and $\Gamma^p$ are just the $T^p$ operators from the Appendix \ref{su21} in the appropriate spaces.
From Eqs. (\ref{ABC}) and
 (\ref{scaling22}) we obtain  the scaling equation as
\begin{eqnarray}
\label{scalingm}
\frac{dJ_A}{d\ln\Lambda}&=&- 2J_A^2- J_B^2\\
\label{scalingm2}
\frac{dJ_B}{d\ln\Lambda}&=&-\frac{3}{2} J_AJ_B- \frac{3}{2} J_BJ_C\\
\label{scalingm3}
\frac{dJ_C}{d\ln\Lambda}&=&-3 J_B^2,
\end{eqnarray}
where the coefficients in the r.h.s. of Eq. (\ref{scalingm}) are: (the number of generators of ${\cal V}_A$ minus one)
times 1 and the number of generators of ${\cal V}_B$
times $\frac{1}{4}$; the coefficients in the r.h.s. of Eq. (\ref{scalingm2}) are: twice the number of generators of ${\cal V}_A$
times $\frac{1}{4}$ and  twice the number of generators of ${\cal V}_C$
times $\frac{3}{4}$; the coefficient in the r.h.s. of Eq. (\ref{scalingm3}) is  the number of generators of ${\cal V}_B$
times $\frac{3}{4}$.

\subsection{Mixing of $U(2)$ ($SU(2)$)  and $U(1)$ symmetries}
\label{partial}

The symmetry of the Hamiltonian considered in the previous Subsection being $U(2)\times U(1)$ or  $U(2)$,  the $su(3)$ algebra was reduced,
to the direct sum of the subspaces (See Appendix \ref{su21}). (Actually, here and further on we should have talked not about the subspaces,
but about the direct products of two  isomorphic subspaces. But we prefer being brief to being rigorous.) Thus the Hamiltonian is block diagonal.

Let us decrease the symmetry of the part  the Hamiltonian acting on the subspace ${\cal V}_A$  to $U(1)$.
Obviously, instead of the single invariant element  of the larger symmetry group on the subspace ${\cal V}_A$ we had previously ($\sum_{p=1}^3\left(T^p\right)^2$),
now there are two invariant elements:
$\sum_{p=1}^2\left(T^p\right)^2$ and  $\left(T^3\right)^2$.
Hence, the most general interaction, satisfying these relaxed symmetry demands, is
\begin{eqnarray}
\label{ABCN}
V&=&2J_{AI}\sum_{p=1}^2G^p\otimes \Gamma^p+2J_{AII}G^3\otimes \Gamma^3 \nonumber\\
&+&2J_{B}\sum_{p=4}^7G^p\otimes \Gamma^p+2J_CG^8\otimes \Gamma^8.
\end{eqnarray}

The scaling equation in this case is
\begin{eqnarray}
\label{scm}
\frac{dJ_{AI}}{d\ln\Lambda}&=&- 2J_{AI}J_{AII}- J_{B}^2\nonumber\\
\frac{dJ_{AII}}{d\ln\Lambda}&=&- 2J_{AI}^2- J_{B}^2\nonumber\\
\frac{dJ_{B}}{d\ln\Lambda}&=&- \left(J_{AI}+\frac{1}{2} J_{AII}\right)J_{B}- \frac{3}{2}J_{B}J_C\nonumber\\
\frac{dJ_C}{d\ln\Lambda}&=&-3J_{B}^2.
\end{eqnarray}
If we put $J_{AI}=J_{AII}=J_A$ we obviously return to Eq. (\ref{scalingm}) -- (\ref{scalingm3}).

\section{Anisotropic Coqblin -- Schrieffer Models}
\label{disc}

\subsection{$XXZ$ Coqblin -- Schrieffer Model}

In our previous paper \cite{kogan2} we introduced a  generalization of the CS model defined by the interaction
\begin{eqnarray}
\label{cs01}
&&V=J_x\sum_{m\neq m'}X_{mm'} c_{m'}^{\dagger}c_m \\
&&+J_z\sum_{m} X_{mm}c_{m}^{\dagger}c_m-\frac{J_z}{N}\sum_{mm'}X_{mm}c_{m'}^{\dagger}c_{m'}.\nonumber
\end{eqnarray}
We called the Hamiltonian the $XXZ$ CS model, because it   was obtained  by  analogy with the $XXZ$ Heisenberg model.
The Hamiltonian turned out to be renormalizable, and scaling equation was
\begin{eqnarray}
\label{scalingsc02}
\frac{dJ_x}{d\ln\Lambda} &=& -(N-2) J_x^2-2J_x J_z\nonumber\\
%\label{scalingsc03}
\frac{d J_z}{d\ln\Lambda} &=& -N J_x^2.
\end{eqnarray}

To connect the previous result with the new ones,
let us  put
$N=3$ in Eq. (\ref{cs01}) and, after a bit of algebra,  write down the equation as
\begin{eqnarray}
\label{ABCd}
V=2J_x{\sum_{p=1}^8}'G^p\otimes \Gamma^p
+2J_z\left(G^3\otimes \Gamma^3+G^8\otimes \Gamma^8\right),
\end{eqnarray}
where  prime means that the terms corresponding to $p=3$ and $p=8$ are excluded from the sum. Note that the operators $T^3$ and $T^8$ generate Cartan subalgebra of the $su(3)$ algebra \cite{pfeifer}.

Equation (\ref{ABCd}) can be easily generalized.  Looking at $N-1$ generators of Cartan subalgebra of the $su(N)$ algebra  (Eq. (\ref{ccdc}) from Appendix \ref{pf})
we realize that the equality
\begin{eqnarray}
\label{hr}
V=2J_x\sum_p'G^p\otimes \Gamma^p+2J_z\sum_qG^q\otimes \Gamma^q,
\end{eqnarray}
where in the first sum  the summation is with respect to $p$ enumerating the generators of the algebra $su(N)$ excluding the generators of Cartan subalgebra, and in the second sum the summation is with respect to the values of $q$ enumerating the generators of Cartan subalgebra   (the latter being given by Eq. (\ref{ccdc})) is valid for any $N$. Thus for $N=4$ there are three terms in the second sum in Eq. (\ref{hr}), corresponding to
$q=3,8,15$ in the notation of generators of Ref. \cite{pfeifer}.

\subsection{$XYZ$ Coqblin -- Schrieffer Model}

In our previous publication  \cite{kogan2} we also introduced a  generalization of the CS model defined by the interaction
\begin{eqnarray}
\label{cs01b}
V&=&\frac{J_x}{2}\sum_{m\neq m'}X_{mm'}\left(c_{m'}^{\dagger}c_{m}+c_{m}^{\dagger}c_{m'}\right)\nonumber\\
&+&\frac{J_y}{2}\sum_{m\neq m'}X_{mm'}\left(c_{m'}^{\dagger}c_{m}-c_{m}^{\dagger}c_{m'}\right)\nonumber\\
&+&J_z\sum_m X_{mm}c_{m}^{\dagger}c_m-\frac{J_z}{N}\sum_{mm'}X_{mm}c_{m'}^{\dagger}c_{m'}.
\end{eqnarray}
We called the model the $XYZ$ CS model, because it   was obtained  by  analogy with the $XYZ$ Heisenberg model.
The Hamiltonian also turned out to be renormalizable, and scaling equation for it was written as
\begin{eqnarray}
\label{scalingsc01b}
\frac{dJ_x}{d\ln\Lambda}&=&-(N-2) J_xJ_y-2J_yJ_z\nonumber\\
%\label{scalingsc02b}
\frac{dJ_y}{d\ln\Lambda}&=&-(N-2) J_xJ_y-2 J_xJ_z\nonumber\\
%\label{scalingsc03b}
\frac{dJ_z}{d\ln\Lambda}&=&-N J_xJ_y.
\end{eqnarray}

The $J_z$ term in the interaction (\ref{cs01b}) is identical to that in Eqs. (\ref{cs01}) and (\ref{hr}). For $N=3$, the interaction, being expressed through the generators, is
\begin{eqnarray}
\label{BCd}
V&=&2J_x\sum_{p=1,4,6}G^p\otimes \Gamma^p+2J_y\sum_{p=2,5,7}G^p\otimes \Gamma^p\nonumber\\
&+&2J_z\left(G^3\otimes \Gamma^3+G^8\otimes \Gamma^8\right).
\end{eqnarray}
Equation (\ref{gm1}) shows that $\lambda_1,\lambda_4,\lambda_6$ are all similar to each other (and similar to $\sigma_x$),
and $\lambda_2,\lambda_5,\lambda_7$ are all similar to each other (and similar to $\sigma_y$). So the similarity of the $XYZ$ CS model (for $N=3$) and the $XYZ$ Kondo model is obvious.
For $N=4$, the interaction (\ref{cs01b}), being expressed through the generators, is
\begin{eqnarray}
\label{BCd2}
V&=&2J_x\sum_{\substack{p=1,4,6,\\9,11,13}}G^p\otimes \Gamma^p+2J_y\sum_{\substack{p=2,5,7,\\10,12,14}}G^p\otimes \Gamma^p\nonumber\\
&+&2J_z\left(G^3\otimes \Gamma^3+G^8\otimes \Gamma^8+G^{15}\otimes \Gamma^{15}\right)
\end{eqnarray}
(again we used the notation of generators of Ref. \cite{pfeifer}).

\section{Discussion}
\label{combi}

One may ask what is the connection between Sections \ref{sym} and \ref{disc}? The answer is that both these Sections follow from Section \ref{se}. More specifically, looking at Eqs. (\ref{ABC}), (\ref{ABCN}), (\ref{hr}), and (\ref{BCd}) we realize that all  renormalizable interactions found by us, using
either the symmetry reasoning, or the analogies with the Heisenberg model,
turn out to be generalizations of Eq. (\ref{productc}). In fact,  all the above mentioned equations can be presented as
\begin{eqnarray}
\label{general}
V=2J_1\sum_{\{p_1\}}G^p\otimes \Gamma^p+2J_2\sum_{\{p_2\}}G^p\otimes \Gamma^p+\dots,
\end{eqnarray}
where $\{p_1\},\{p_2\}, \dots$ present the partition of all the generators of the appropriate ($su(3)$ for Eqs. (\ref{ABC}), (\ref{ABCN}),  and (\ref{BCd}), $su(N)$ for Eq.
(\ref{hr}))  Lie algebra.
It would be interesting to try to look at the general Eq. (\ref{general}) from the point of view of the theory of invariants.

\section{Conclusions}
\label{conclusions}

We studied algebraic structure of the scaling equation for a general model, describing   a  quantum impurity with degenerate energy levels, interacting with a gas of itinerant electrons.
More specifically we studied the connection between the explicit form of the scaling equation and the symmetry of interaction.
On the basis of this analysis
we have written down  the scaling equations  for the Hamiltonians having $U(2)$ ($SU(2)$) symmetry.
We also presented a new representation for the anisotropic CS models proposed by us earlier.

\begin{acknowledgments}

The author is grateful to N. Andrei, Y. Avishai, H. E. Haber,   V. Yu. Irkhin, J. Kroha, M. Pletyukhov, S. Shallcross, Zh. Shi, A. Sinner, A.  Weichselbaum, O. M. Yevtushenko, V. Yudson, and P. Zalom  for valuable discussions.

The work on this paper  started during the author's  visit to
Max-Planck-Institut fur Physik komplexer Systeme in 2019.
The paper was finalized during the author's visit to
DIPC, San Sebastian/Donostia.
The author  cordially thanks both Institutions for the hospitality extended to him during
those and all his  previous visits.

\end{acknowledgments}

\begin{appendix}

\section{Maximally symmetric interaction analysed  in mutually dual bases}
\label{curved}

In Section \ref{se} we   assumed that the   generators are self-dual.
However, the results may be written down in the basis independent form.
In fact,
analysing   Eq. (\ref{product})  we realise that for the matrices $\Gamma$ we can use the basis
 different from that used for the matrices $G$. Distinguishing the former by the sign tilde above the generators we can rewrite Eqs.(\ref{product}) and (\ref{scaling2}) respectively as
\begin{eqnarray}
\label{prb}
V=2\sum_{pp'}c_{pp'}G^p\otimes \widetilde{\Gamma}^{p'}
\end{eqnarray}
and
\begin{eqnarray}
\label{sca2}
\frac{dc_{pp'}}{d\ln\Lambda}=-\sum_{sts't'}f^p_{st}\widetilde{f}^{p'}_{s't'}c_{ss'}c_{tt'}.
\end{eqnarray}

If we chose the tilde basis as the dual basis to that without the tilde, whatever the former is, the
invariant element of the algebra can be written in the form \cite{humphreys}
\begin{eqnarray}
\label{prodc}
\text{inv}=\sum_{p}G^p\otimes \widetilde{\Gamma}^{p},
\end{eqnarray}
and Eq. (\ref{sc}) becomes
\begin{eqnarray}
\sum_{st}f^p_{st}\widetilde{f}^{p'}_{st}=N \delta_{p,p'}.
\end{eqnarray}
Hence   the interaction having the full symmetry $G$  is
\begin{eqnarray}
\label{pructc}
V=2J\sum_{p}G^p\otimes \widetilde{\Gamma}^{p},
\end{eqnarray}
and we regain Eqs. (\ref{scaling4}), but this time written down in an arbitrary basis.

\section{Non-symmetric $XXZ$ Kondo model}
\label{DM}

Everywhere in the main body of the paper  we assume that the interaction  matrix $c_{pp'}$  is symmetric.
Here we want to show, using the $XXZ$ Kondo model as an example, what the removal of this limitation can lead to.
In distinction to the treatment of the model in the Subsection \ref{guide}, there appears now an additional invariant:  $S^x\otimes \sigma^y-S^y\otimes \sigma^x$, and
the most general interaction with the symmetry $U(1)$ can be written as an arbitrary linear combination of the three invariant elements
\begin{eqnarray}
\label{aab}
V&=&J_x\left(S^x\otimes \sigma^x+S^y\otimes \sigma^y\right)+J_zS^z\otimes \sigma^z\nonumber\\
&+&J_{DM}\left(S^x\otimes \sigma^y-S^y\otimes \sigma^x\right)
\end{eqnarray}
(the index $DM$ standing for Dzyaloshinskii-Moriya).
Using Eq. (\ref{scaling2}), we get   the scaling equation  as
\begin{eqnarray}
\label{scaling45}
\frac{dJ_x}{d\ln\Lambda}&=&- 2J_xJ_z\nonumber\\
\frac{dJ_{DM}}{d\ln\Lambda}&=&- 2J_{DM}J_z\nonumber\\
\frac{dJ_z}{d\ln\Lambda}&=&- 2J_x^2-2J_{DM}^2.
\end{eqnarray}
The two first integrals of Eq. (\ref{scaling45}) are obvious
\begin{eqnarray}
J_z^2-J_x^2-J_{DM}^2=\text{const},\;\;\;
\frac{J_{DM}}{J_x}=\text{const}.
\end{eqnarray}
So the equation can be easily integrated in terms of trigonometric (hyperbolic) functions.

\section{Group $SU(3)$ and its subgroups}
\label{su21}

The  generators  of $su(3)$ algebra can be chosen as
$T^p=\lambda^p/2$,
where  the Gell-Mann matrices
\begin{eqnarray}
\label{gm1}
\lambda^1&=&\left(\begin{array}{ccc} 0 & 1 & 0 \\ 1 & 0 & 0\\ 0 & 0 & 0\end{array}\right),
\lambda^2=\left(\begin{array}{ccc} 0 & -i & 0 \\ i & 0 & 0\\ 0 & 0 & 0\end{array}\right),
\lambda^3=\left(\begin{array}{ccc} 1 & 0 & 0 \\ 0 & -1 & 0\\ 0 & 0 & 0\end{array}\right),\nonumber\\
\lambda^4&=&\left(\begin{array}{ccc} 0 & 0 & 1 \\ 0 & 0 & 0\\ 1 & 0 & 0\end{array}\right),
\lambda^5=\left(\begin{array}{ccc} 0 & 0 & -i \\ 0 & 0 & 0\\ i & 0 & 0\end{array}\right), \nonumber\\
\lambda^6&=&\left(\begin{array}{ccc} 0 & 0 & 0 \\ 0 & 0 & 1\\ 0 & 1 & 0\end{array}\right),
\lambda^7=\left(\begin{array}{ccc} 0 & 0 & 0 \\ 0 & 0 & -i\\ 0 & i & 0\end{array}\right),
\end{eqnarray}
and
\begin{eqnarray}
\label{gm2}
\lambda^8&=&\frac{1}{\sqrt{3}}\left(\begin{array}{ccc} 1 & 0 & 0 \\ 0 & 1 & 0\\ 0 & 0 &-2\end{array}\right)
\end{eqnarray}
are the $su(3)$ analogs of the Pauli matrices \cite{georgi}.
With this choice of generators, the structure constants $f^p_{st}$ are totally antisymmetric with respect to the interchange of any
pair of indices, and are given by \cite{georgi}
\begin{eqnarray}
\label{commutator1}
f_{12}^3&=&1,\\
\label{commutator2}
f_{14}^7&=&f_{16}^5=f_{24}^6=f_{25}^7=f_{34}^5=f_{37}^6=\frac{1}{2},\\
\label{commutator3}
f_{84}^5&=&f_{86}^7=\frac{\sqrt{3}}{2},
\end{eqnarray}
while all other $f_{st}^p$ not related to these by permutation are zero.  It will be important for us in Subsection \ref{suu} that there are only three different squares of structure constants: $1$, $\frac{1}{4}$ and  $\frac{3}{4}$.

The group $SU(3)$ acts adjointly on the $su(3)$ algebra.
The matrix elements of
the generators of the adjoint representation $D$ are  \cite{jones}
\begin{eqnarray}
\label{j}
\left(D(T^p)\right)^s_t\equiv 2\text{Tr}\left(T^s[T^p,T^t]\right)=if^s_{pt},
\end{eqnarray}
The algebra $su(3)$ realises irreducible representation of the group $SU(3)$. Consider now the adjoint action of the  subgroup $SU(2)$,
generated by $[T^1,T^2,T^3]$, on the same algebra. In this case
the $su(3)$ algebra, considered as just a vector space, can be reduced to a   direct sum of three subspaces:
${\cal V}_A$ with three generators $[T^1,T^2,T^3]$, ${\cal V}_B$ with four generators $[T^4,T^5,T^6,T^7]$, and ${\cal V}_C$ with a single generator $T^8$
\begin{eqnarray}
\label{separ}
su(3)\to {\cal V}_A\oplus {\cal V}_B\oplus {\cal V}_C.
\end{eqnarray}
Looking at Eqs. (\ref{commutator1}) -(\ref{commutator3}) we realise that each subspace  is closed under the adjoint action of $SU(2)$,
thus we have decomposed the representation of the group $SU(2)$, realised on the algebra $su(3)$, into three representations.

For each of the subspaces there obviously exist an invariant  (with respect to algebra generated by $[T^1,T^2,T^3]$) element, which is equal to the sum of all the subspace generators.
In each case it is the only such element, because the representations realised on each of subspace are irreducible.

Now consider  the adjoint action of the group  $U(2)=SU(2)\times U(1)$ on the  algebra $su(3)$.
The group has additional, in comparison with the group $SU(2)$, symmetry generator $T^8$, which commutes with each of the  generators $[T^1,T^2,T^3]$. \cite{elliott}
Because the generator $T^8$ do not mix the subspaces introduced above, the representation decomposition  is still valid. Also, because all the representations considered above were irreducible for the group  $SU(2)$,
they remain as such after the upgrading the symmetry. One can check up that all the previous invariant elements are the invariant elements of $T^8$ also.
So finally, both the group $SU(2)\times U(1)$ and the group $U(2)$, acting on the algebra $su(3)$ has three and only three quadratic invariant elements: $\sum_{p=1}^3\left(T^p\right)^2$,
$\sum_{p=4}^7\left(T^p\right)^2$, and $\left(T^8\right)^2$.

\section{Generators of Cartan subalgebra of the $su(N)$ algebra}
\label{pf}

$N-1$ generators of Cartan subalgebra of the $su(N)$ algebra are \cite{pfeifer}
\begin{eqnarray}
\label{ccdc}
&&\frac{1}{2}\left(\begin{array}{ccccccc} 1 &  &  & &&&\\  & -1 &&&&&\\ &&0 & &&& \\ &&&0  &&&\\ && && \cdot & & \\ &&& & &\cdot  &\\&&&&&&0
\end{array}\right),
\frac{1}{2\sqrt{3}}\left(\begin{array}{ccccccc} 1 &  &  & &&&\\  & 1 &&&&&\\ &&-2 & &&& \\ &&&0  &&&\\ && && \cdot & & \\ &&& & &\cdot  &\\&&&&&&0
\end{array}\right),\nonumber\\
&&\dots,\frac{1}{\sqrt{2N(N-1)}}\left(\begin{array}{ccccccc} 1 &  &  & &&&\\  & 1 &&&&&\\ &&1 & &&& \\ &&&\cdot  &&&\\ && && \cdot & & \\ &&& & &1  &\\&&&&&&-N+1
\end{array}\right)
\end{eqnarray}

\end{appendix}


\begin{thebibliography}{99}


\bibitem{kondo} J. Kondo,  Prog. Theor. Phys. {\bf 32}, 37 (1964).

\bibitem{hewson} A. C. Hewson, {\it The Kondo Problem to Heavy Fermions}, (Cambridge University Press, Cambridge, 1993).

\bibitem{anderson} P. W. Anderson, J. Phys. C {\bf 3}, 2436 (1970).

\bibitem{noz} Ph. Nozieres and A. Blandin, Journal de Physique, {\bf 41},  193 (1980).

\bibitem{kroha2} M. Arnold, T. Langenbruch, and J. Kroha, \prl {\bf 99}, 186601 (2007)

\bibitem{cox} D. L. Cox and A. Zawadowski, Adv. Phys. {\bf 47},  599 (1998).

\bibitem{kroha} A. Rosch, J. Paaske, J. Kroha, and P. Wolfle,
Phys. Rev. Lett. {\bf 90}, 076804 (2003).

\bibitem{koerting}  V. Koerting, P. Fritsch, and  S. Kehrein, Physica B: Condensed Matter
{\bf 406}, 2091 (2011).

\bibitem{kikoin} A. K. Kikoin, M. Kiselev, Y. Avishai, {\it Dynamical Symmetries
for Nanostructures} (Springer-Verlag/Wien, 2012).

\bibitem{krones} J. Krones and G. S. Uhrig, Phys. Rev. B {\bf 91}, 125102 (2015).

\bibitem{florens} S. Florens and I. Snyman, Phys. Rev. B 92, 195106 (2015).

\bibitem{delft}  C. J. Lindner, F. B. Kugler, H. Schoeller, J. von Delft, Phys. Rev. B {\bf 97}, 235450 (2018).

\bibitem{zalom}  P. Zalom, J. de Bruijckere, R. Gaudenzi, H. S. J. van der Zant, T. Novotny, R. Korytar,  arXiv:1901.08514 (J. Phys. Chem. C, accepted for publication).

\bibitem{shiba} H. Shiba, Prog.  Theor. Phys. {\bf 43}, 601 (1970).

\bibitem{yosida} K. Yosida, {\it Theory of Magnetism} (Springer, Berlin Heidelberg New York, 1996).

\bibitem{irkhin2} V. Yu. Irkhin and Yu. P. Irkhin, JETP {\bf 107}, 616 (1995);
  V. Yu. Irkhin and M. I. Katsnelson,
 Phys. Rev. B {\bf 59}, 9348 (1999).

\bibitem{costi} G. Zarand, T. Costi, A. Jerez, and N. Andrei, Phys. Rev. B {\bf 65}, 134416 (2002).

\bibitem{thomas} C. Thomas,  A. S. da Rosa Simoes, C.  Lacroix, J.  R. Iglesias,  B. Coqblin,
Physica B, {\bf 404}, 3008 (2009).

\bibitem{kogan} E. Kogan, K. Noda, and S. Yunoki,
Phys. Rev. B {\bf 95}, 165412 (2017).

\bibitem{kogan2} E. Kogan,  J. Phys. Commun. {\bf 2}, 085001 (2018).

\bibitem{kopietz} D. Tarasevych, J. Krieg, and P. Kopietz, Phys. Rev. B {\bf 98}, 235133 (2018).

\bibitem{coqblin} B. Coqblin, J.R. Schrieffer, Phys. Rev. {\bf 185}, 847 (1969).

\bibitem{rajan} V. T. Rajan, Phys. Rev. Lett. {\bf 51}, 308 (1983).

\bibitem{schlottmann} P. Schlottmann, Zeitschrift fur Physik B: Condensed Matter {\bf 51}, 223 (1983).

%\bibitem{kashiba} Shin-ichi Kashiba, S. Maekawa, S. Takahashi, and Masashi Tachiki
%J. Phys. Soc. Jpn. {\bf 55},  1341 (1986).

\bibitem{andrei}  A. Jerez, N. Andrei, and G. Zarand, Phys.
Rev. B {\bf 58}, 3814 (1998).

\bibitem{zlatic} V. Zlatic, B. Horvatic, I. Milat, B. Coqblin, G. Czycholl, and C. Grenzebach, Phys. Rev. B {\bf 68}, 104432
(2003).

\bibitem{bazhanov}  V. V. Bazhanov, S.L. Lukyanov, A.M. Tsvelik, Phys. Rev. B  {\bf 68}, 094427 (2003).

%\bibitem{kuzmenko} I. Kuzmenko, Y. Avishai, Phys. Rev. B {\bf 89}, 195110 (2014).

\bibitem{desgranges} H.-U. Desgranges, Physica B: Condensed Matter {\bf 454}, 135 (2014); {\bf 473}, 93 (2015).

\bibitem{figueira} M. S. Figueira, A. Saguia, M. E. Foglio, J. Silva-Valencia, and R Franco, Physica B: Condensed Matter
{\bf 455}, 92 (2014)

\bibitem{avishai}  I. Kuzmenko, T. Kuzmenko, Y. Avishai, and Gyu-Boong Jo, Phys. Rev. B   {\bf 97}, 075124, (2018).

\bibitem {hall} B. C. Hall, {\it Lie Groups, Lie Algebras, and Representations: An Elementary Introduction, Graduate Texts in Mathematics, 222 (2nd ed.)}, (Springer, 2015).


\bibitem{georgi} H. Georgi, {\it Lie Algebras in Particle Physics (2nd ed.)}, (Taylor \& Frencis, 2018).

\bibitem{elliott} J. P. Elliott and P. G. Dawber, {\it Symmetry in Physics, vol. 1:
Principles and Simple Applications}, (Macmillan Press LTD, 1979).

\bibitem{pfeifer} W. Pfeifer, {\it The Lie algebras $su(N)$: an introduction}, (Birkhauser, Basel-Boston-Berlin, 2003)).

\bibitem{humphreys} J. E. Humphreys,  {\it Introduction to Lie Algebras and Representation Theory. Graduate Texts in Mathematics.  (Second printing, revised ed.)}, (Springer-Verlag,  New York,  1978).

\bibitem{jones} H. F. Jones, {\it Groups, Representations and Physics, (2 ed.)}, (Institute of Physics Publishing, Bristol and Philadelphia (2003).



\end{thebibliography}
\end{document}